\documentclass[a4paper,12pt]{article}

\usepackage{amsmath,amssymb,array,amsfonts}
\usepackage[dvipdfmx]{graphicx}
\usepackage{comment,color}
\usepackage{geometry}
\usepackage[bf, small, hang]{caption}
\allowdisplaybreaks

\newcommand{\be}{\begin{equation}}
\newcommand{\ee}{\end{equation}}
\newcommand{\bea}{\begin{eqnarray}}
\newcommand{\eea}{\end{eqnarray}}

\makeatletter
    
    \@addtoreset{equation}{section}
\makeatother

\usepackage
[colorlinks=true,
 linkcolor=black,
 urlcolor=magenta,
 citecolor=cyan,
 hypertex]
 {hyperref}
\begin{document}
\setlength{\baselineskip}{18pt}
\begin{titlepage}

\begin{flushright}
KOBE-TH-14-07%
%HRI-P-12-01-001 
\end{flushright}
\vspace{1.0cm}
\begin{center}
{\Large\bf Anomalous Higgs interactions \\
\vspace*{3mm}
in dimensional deconstruction} 
\end{center}
\vspace{25mm}

\begin{center}
{\large
Nobuaki Kurahashi, C.S. Lim$^*$ and Kazuya Tanabe
}
\end{center}
\vspace{1cm}
\centerline{{\it
Department of Physics, 
Kobe University, Kobe 657-8501, Japan}} 

\centerline{{\it
$^*$The Department of Mathematics, Tokyo Woman's Christian University, Tokyo 167-8585, Japan}}

%
%%%%%%%%%%%%%%%%%%%%%%%%%%%%%%%%%%%%%%%%%%%%%%%%%%%%%%%%%%
%%          %%%%%%%%%%%%%%%%%%%%%%%%%%%%%%%%%%%%%%%%%%%%%%
%% Abstract %%%%%%%%%%%%%%%%%%%%%%%%%%%%%%%%%%%%%%%%%%%%%%
%%          %%%%%%%%%%%%%%%%%%%%%%%%%%%%%%%%%%%%%%%%%%%%%%
%%%%%%%%%%%%%%%%%%%%%%%%%%%%%%%%%%%%%%%%%%%%%%%%%%%%%%%%%%
%
\vspace{2cm}
\centerline{\large\bf Abstract}
\vspace{0.5cm}

Recent LHC experiments have revealed that Higgs is light. As an interesting candidate to accommodate light Higgs, in this paper we adopt the scenario of dimensional deconstruction, where Higgs is redarded as a pseudo-Nambu-Goldstone boson. Though the scenario is formulated in ordinary 4-dimensional space-time, it may also be interpreted as ``latticized" gauge-Higgs unification. We point out that in this scenario Higgs interaction with matter field is anomalous, i.e. its coupling deviates from what the standard model predicts. The interplay between the periodicity of physical observables in the Higgs field and the violation of translational invariance along the extra-space due to the latticization is argued to play an essential role to get the anomalous interaction. Though the predicted anomalous Higgs interaction has much similarity to the one in the gauge-Higgs unification, in the case of dimensional deconstruction the anomaly exists even if we do not introduce bulk mass term for the chiral fermion realized by orbifolding, in clear contrast to the case of gauge-Higgs unification. It in turn means that the anomaly goes away in the continuum limit of the extra-space.

\end{titlepage}

%\maketitle
%\thispagestyle{empty}

%\vspace{5mm}

%}

\newpage
%%%%%%%%%%%%%%%%%%%%%%%%%%%%%%%%%%%%%%%%%%%%%%%%%%%%%%%%%%
%%              %%%%%%%%%%%%%%%%%%%%%%%%%%%%%%%%%%%%%%%%%%
%% Introduction %%%%%%%%%%%%%%%%%%%%%%%%%%%%%%%%%%%%%%%%%%
%%              %%%%%%%%%%%%%%%%%%%%%%%%%%%%%%%%%%%%%%%%%%
%%%%%%%%%%%%%%%%%%%%%%%%%%%%%%%%%%%%%%%%%%%%%%%%%%%%%%%%%%
\section{Introduction} 

In spite of the great success of the LHC experiments \cite{ATLAS}, \cite{CMS} to have discovered the Higgs particle, we still do not know whether the discovered scalar particle is what we expect in the standard model or a particle some theory of physics beyond the standard model (BSM) has in its low energy effective theory. In other words, we do not have any conclusive argument of the origin of the Higgs. 

The experiments, however, have provided us with very important information of the Higgs particle: the observed Higgs mass $M_{H} = 126$ GeV is of the order of the weak scale $M_{W}$. Namely, the Higgs has turned out to be ``light", which suggests that the quartic   self-coupling of the Higgs $\lambda$ is of ${\cal O}(g^{2})$ ($g$: gauge copling constant) and, therefore, is handled by gauge principle. 

Among possible BSM models, we can pick up a few candidates where the Higgs self-coupling is handled by gauge interactions. First candidate  is MSSM (Minimal Supersymetric Standard Model), where the coupling $\lambda$ comes only from the D-term contribution, leading to $M_{H} \leq \cos 2\beta M_{Z}$ at the classical level, though a sizable quantum correction is expected to explain the observed Higgs mass assuming larger SUSY breaking mass scale.  

There is another candidate of BSM: Gauge-Higgs Unification (GHU) \cite{1979Manton}, \cite{1983Hosotani}. In this scenario, Higgs field is identified with the extra-space component of the higher-dimensional gauge field. For instance, in the simplest 5-dimensional (5D) case, the higher-dimensional gauge field is decomposed as 
\be 
\label{1.1} 
A_{M} = (A_{\mu}, A_{y}) \ \ \ (\mu = 0,1,2,3), 
\ee 
where $A_{\mu}$ just behaves as ordinary 4D gauge field, while (the Kaluza-Klein (KK) zero mode of) the extra-space component $A_{y}$ is identified as the Higgs field. The fact that the Higgs is originally a gauge boson in GHU provides a new type of solution for the hierarchy problem relying on higher dimensional gauge symmetry, thus opening a new avenue for BSM theories \cite{1998HIL}. In this scenario, the Higgs self-coupling is handled by gauge princile, just because Higgs is originally a gauge field. Interestingly, in a 6D GHU on a two dimensional orbifold $T^{2}/Z_{3}$ an attractive relation $M_{H} = 2M_{W}$ holds at the classical level \cite{SSSW} and it has been pointed out that the quantum correction to this relation is calculable as a UV-finite value \cite{2014LMM}, just as in the case of MSSM.  

It is worth noticing that both of MSSM and GHU have been proposed aiming to solve the hierarchy problem: the problem of quadratically  divergent quantum correction to the Higgs mass. There also exist other types of well-discussed BSM theories formulated in 4D space-time, proposed in order to solve the hirerarchy problem, which  have close relationship with GHU. Namely, the scenarios of \\ 
(i) Dimensional deconstruction (DD) \cite{2001ACG}, \\ 
(ii) Little Higgs (LH) \cite{Schmaltz}. \\ 
Basically, in both of these scenarios Higgs is regarded as pseudo-Nambu-Goldstone boson (PNGB) and therefore is non-linearly realized in a form $e^{i \frac{\phi}{f}}$, where $\phi$ denotes the PNGB treated as the Higgs field and $f$ corresponds to its ``decay constant". Let us note that this leads to an important consequence that phyiscal observable are periodic in the Higgs field $\phi$ in these scenarios. Similar periodicity exists, as we will discuss below, in the scenario of GHU as well. As long as the global symmetry is exact the Higgs mass vanishes, which suggests a light Higgs. 

As another circumstancial evidence of the mutual relation between GHU and DD or LH, it may be pointed out that both have shift symmeties, i.e. the invariance under the transformations
\be
\label{1.2} 
A_{y} \ \to A_{y} + \partial_{y} \lambda \ \ (\mbox{for GHU}), \ \ \ \phi \ \to \phi + c \ \ (\mbox{for DD, LH}),  
\ee 
where the former is nothing but higher dimensional gauge transformation (in the simplest U(1) 5D GHU), and the latter is some global tramsformation with parameter $c$, like a phase transformtion in the simplest U(1) global symmetry. These symmetries strictly forbid the presence of local operators responsible for the masses of $A_{y}$ and $\phi$, 
thus solving the hierarchy problem at quantum level. 

The mutual close relation becomes manifest and more solid, once we realize that the DD scenario may be regarded as a sort of GHU, where the extra space is latticized to several lattice points and $e^{i \frac{\phi}{f}}$ is regarded as the product of ``link variables" (or Wilson-loop) in a lattice gauge theory, as we will discuss later. It's also worth while noting that the scenario of the LH was inspired by the DD scenario. 

After the discovery of the Higgs, now the main focus of particle physics may be on checking whether its interactions are what the SM predicts or not. If some deviations from the predictions made by the SM are found by precision tests of the Higgs interactions it will clearly signal the presence of new physics. The precision tests of the Higgs interactions should be one of the main purposes of the proposed ILC experiment. 

In this paper we discuss such ``anomalous Higgs interactions", i.e. the Higgs couplings which deviate  from what the SM predicts. 

All of BSM scenarios mentioned above extend the Higgs sector or give new interpretations of the origin of the Higgs, since these were deviced in order to solve the hierarchy problem of the Higgs sector relying on some symmetries. Thus it may not be surprising even if these scenarios predict anomalous Higgs interactions. 

As the matter of fact, in the previous paper \cite{2013HKLT} we have seen that in the scenario of GHU, as its characteristic prediction, anomalous Yukawa couplings are predicted, as was originally pointed out in \cite{2007HS}, \cite{2008HOOS}, \cite{2009HK}, \cite{2009HKT}.

Probably one of the most important observation in our analysis concerning the anomalous intrecations in the scenario of GHU is that the anomaly is inevitable consequence of the following two properties of the theory: \\ 
(a) The periodicity of the physical observables in the Higgs field,  \\
(b) The violation of the translational invariance along the extra space. \\ 
Let us discuss a little why these properties are essential to get the anomaly and how they appear in the GHU scenario. 

First, concerning the issue (a), we should note that in GHU, the Higgs field may be interpreted as a Wilson-loop phase or a sort of AB (Aharonov-Bohm) phase, at least in 5D space-time where the extra space is a circle, a non-simply connected space. Namely, the circle allows the penetration of magnetic flux inside itself and the zero-mode of $A_{y}$ is regarded as a vector potential generated by the magnetic flux. Thus the VEV of Higgs, though it is just a constant, is not a pure-gauge and cannot be gauged away. On the other hand, as long as the Higgs field appears through a phase, we naturally expect that physical observables in this theory are all periodic in the Higgs field. 

Typical periodic functions are trigonometric functions. In fact, for the KK zero-modes of lighter quarks (say, the quarks of first and second generations), their masses behave as sine-functions of the VEV of the Higgs $v$ \cite{2013HKLT}, \cite{2007HS}, \cite{2008HOOS}, \cite{2009HK}, \cite{2009HKT}: 
\be 
\label{1.3}
m(v) \propto \sin (\frac{g_{4}}{2}\pi R v),  
\ee
where $g_{4}$ is the 4D gauge coupling and $R$ is the radius of the circle, the extra space. Since Higgs interaction is expected to be obtained by the replacement $v \to v+h$ with $h$ being physical Higgs field, the Yukawa coupling (to be precise, the ``diagonal part" of the Yukawa coupling matrix in the base of KK modes for each flavor of quarks) is given by the first derivative of the ``mass function" $m(v)$: 
\be 
\label{1.4}
f = \frac{dm(v)}{dv} \propto \cos (\frac{g_{4}}{2}\pi R v), 
\ee
which is no longer constant as in the standard model (SM), since the mass function is not linear in $v$. Thus the Yukawa coupling for light quarks predicted by GHU, $f_{GHU}$, deviates from that in the SM, $f_{SM}$, as \cite{2013HKLT} 
\be 
\label{1.5} 
\frac{f_{GHU}}{f_{SM}} \simeq x \cot x \ \ \ (x \equiv \frac{g_{4}}{2}\pi R v), 
\ee 
which even vanishes for a specific value of the VEV, $x = \frac{v}{2}$ \cite{2007HS}, \cite{2008HOOS}, \cite{2009HK}, \cite{2009HKT}, though the ratio reasonably approaches to unity for small $x$, since it this limit $M_{W} \ll \frac{1}{R}$ and the SM is expected to be recovered as the low energy effective theory of GHU. 

One remark is in order here. In a realistic GHU model, to make the theory chiral, we adopt an orbifold as the extra space. In the case of 5D GHU, the orbifold is $S^{1}/Z_{2}$. In this framework, ``$Z_{2}$-odd bulk mass term" 
\be 
\label{1.6}
\epsilon (y) M \bar{\psi}\psi 
\ee 
is allowed, being $Z_{2}$ invariant, where $\epsilon (y)$ is the sign function of the extra-space coordinate $y$: $\epsilon (y) = \pm 1$ depending on the sign of $y$. This mass term causes the localization of mode functions of chiral fermions at different fixed points of the orbifold depending on their chiralities and the overlap of their mode functions is exponentially suppressed, thus naturally leading to hierarchically small (zero-mode) quark masses for lighter generations. The sine-function (\ref{1.3}) is known to be obtained as the result of the presence of this  bulk mass term as we discuss below. Let us note that this $Z_{2}$-odd bulk mass term, behaving as a sort of kink-solution of scalar field, clearly violates the translational invariance along the extra-space.     

The argument above seems to suggest that just the periodicity leads to the anomalous Yukawa coupling. Then what role does the property or condition (b) play to get the anomaly ? Actually, the periodicity (a) does not necessarily lead to such non-linear mass function as in (\ref{1.3}). In fact, in \cite{2013HKLT} we have argued that for the KK zero-mode of heavy quark like top quark, the mass function is just linear function of $v$, just as in the SM, while the periodicity is still guaranteed by the replacement of KK zero-mode by first KK mode at the ``crossing point" of mass spectrum at $x = \frac{\pi}{2}$. The essence of the argument is that for a heavy quark the $Z_{2}$-odd bulk mass $M$ can be switched off and the translational invariance along the extra-space is not violated (to be more precise, the absolute values of extra-space momentum or KK modes are preserved even under the presence of fixed points), and the mass function for zero-mode is ``normal" linear function of $v$, just as in the orginal argument in the case of $S^{1}$ compactification \cite{1998HIL}. At the crossing point KK zero mode and first KK mode do not mix each other due to the translational invariance, i.e. the conservation of extra-space momentum $p_{y}$.

What happens if we switch on a small bulk mass $M$ ? Now the mixing between zero and first KK modes arises and the degeneracy of mass spectrum at the level crossing is lifted a little and linear function is slightly modified,  accordingly. When the bulk mass becomes sizable, as in the case of lighter quarks, the mode function of zero mode is considerably modified and we finally obtain the sine-function as shown in (\ref{1.3}). We thus realize that anomalous interaction is the inevitable consequence of the interplay of the periodicity and the violation of the translational invariance, as mentioned in (a) and (b) above. 

The purpose of this paper is to discuss anomalous Higgs interaction in the scenario of 
dimensional deconstruction (DD) \cite{2001ACG}. 

As has been already mentioned above, in the original proposal of DD scenario \cite{2001ACG} the Higgs is a pseudo-Nambu-Goldstone boson (PNGB) composed by a pair of fermion and anti-fermion by strong interactions, just as the pions are composed by pairs of quarks and anti-quarks in QCD. The specific feature of the model 
is that after the confinement by the strong interactions the remaining (weak) gauge symmetries are of the form of the direct product of the same type of gauge group: SU$_1$(m) $\times$ SU$_2$(m)$\times \cdots \times$ SU$_N$(m), as is schematically displayed by the ``moose diagram". 

From a different point of view, the DD scenario can be interpreted as a sort of GHU scenario, where the extra-space is ``latticized", $N$ being the number of the lattice sites. In fact, the non-linear realization of the Higgs field $\phi$
\be 
\label{DD1.1} 
U = e^{i\sqrt{N}\frac{\phi}{f}}
\ee
just corresponds to the Wilson-loop in GHU. 

Such close relation of DD with GHU strongly suggests that similar anomalous Higgs interactions to those in GHU are expected in DD scenario as well. We show in this paper that it is raeally the case. 

We easily understand that the conditions (a) and (b) to get anomalous interaction mentioned above are met also in DD scenario, we are now interested in. First, the condition (a), i.e. the periodicity in the Higgs field also exists in this theory, since the Higgs field is non-linearly realized as a sort of phase factor as is seen in (\ref{DD1.1}). Secondly, it is clear that the translational invariance is violated by the fact that the extra-space is latticized, once DD is understood as a latticized 5D GHU. Thus it is almost promissing that we get anomalous Higgs interactions in the scenario of DD. 

We, however, also should note that there is a qualitatively distinct difference in the anomalous interaction present in DD scenario from the one in the scenario of GHU. Namely, it is because of the violation of the translational invariance by latticizing the extra-space, not because of the $Z_{2}$-odd bulk mass for fermion as in the case of GHU. As far as the violation is due to the property of the space-time itself on which the theory is constructed, the anomalous interactions should arise not only in the sector of matter fermion but basically in every sector of the theory. The situation, in such a sense, may be similar to the case of GHU formulated on the Randall-Sundrum type 5D space-time, where the translational invariance is violated by the presence of the ``warp-factor" $e^{- \kappa |y|}$ and anomalous interactions appear not only in the fermionic sector but also in the gauge boson sector as well \cite{2007HS}. This on the other hand suggests that the anomaly goes away in the limit $a \to 0$ ($a$: lattice spacing), unless there is no other source of violation of translational invariance.   

In this paper, we take an attitude that the snenario of DD is equivalent to a latticized GHU and study the anomalous Higgs interaction in the GHU with latticized extra-space. What we adopt as a model is 5D scalar QED wtih latticized extra-space as its compactified extra-space \cite{2003KSS}.

After all, Higgs interactions are devided into two categories, i.e. anomalous versus normal, and to see which categorie is chosen by the Nature is quite important in order to conclude whether physics BSM is realized or not and, if it is ever realized, which type of BSM is selected.

\section{5D gauge theory with latticized extra-space} 

Before discussing 5D QED, here we consider a generic 5-dimensional (5D) gauge theory where the extra dimension is compactified $S^1$ of radius $R$ and circumference $L = 2\pi R$, which is latticized to $N$ littice sites with extra space coordinates $y_{i} \ (i = 1 - N)$. 

For a generic matter field $\psi (x^{\mu}, y_{i})$, a local gauge transformation is given as 
\be 
\label{DD2.1} 
\psi (x^{\mu}, y_{i}) \ \ \to \ \ \psi'(x^{\mu},y_j) = g(x^{\mu},y_j)\psi(x^{\mu},y_j), 
\ee
where $g(x^{\mu},y_j)$ is a member of the gauge group $\mathcal{G}$. In the limit of $N \to \infty$ this trasnformation reduces to a 5D gauge transformation. On the other hand, we may regard it as the 4D local gauge transformation whose gauge group is a direct product:  
\be 
\label{DD2.2}
G_{1} \times G_{2} \times \cdots \times G_{N} 
\ee
where each of $G_{j}$, with the group element $g(x^{\mu},y_j)$, belongs to the same group $\mathcal{G}$. 
(\ref{DD2.2}) is just equivalent to the the gauge symmetry shown by the ``moose diagram" in the original DD scenario \cite{2001ACG}, where $G_{i}$ are ``weak" gauge symmetries remaining after the confinement due to the strong forces. In this way, we can confirm that the scenario of DD is equivalent to the GHU where the extra-space is latticized. 

Hereafter, we change the notation of the field as 
\be 
\label{DD2.3} 
\psi(x^{\mu},y_j) \ \ \to \ \ \psi_{i}(x^{\mu}).  
\ee
The fields $\psi_{i}(x^{\mu}) \ (i = 1, 2, \ldots, N)$ may also be regarded as $N$ pieces of 4D fields. 
Because of the $S^1$ compactification, there is a periodic boundary conditions as follows,
\begin{equation} 
\label{DD2.4}
\psi_{N+i}(x^{\mu}) = \psi_{i}(x^{\mu}).
\end{equation}

The ``derivative" along the extra-space is given by a difference,
\begin{equation}
\label{DD2.5} 
\partial_{y}\psi_{i}(x^{\mu}) \equiv \frac{\psi_{i+1}(x^{\mu}) - \psi_{i}(x^{\mu})}{a},
\end{equation} 
where $a$($>0$) is a distance between neighboring sites: lattice spacing, satisfying 
\be 
\label{DD2.6} 
L \equiv 2\pi R = Na. 
\ee 

The covariant derivative along 4D space-time is just an ordinary one 
\be 
\label{DD2.7}
D_{\mu} \psi_{i} = \partial_{\mu}\psi_{i} - igA_{i \mu}\psi_{i}.
\ee
where $A_{i \mu}$ is the gauge field of $G_{i}$ and $g$ is the gauge coupling constant.

Covariant derivative along extra dimension is given by  
\be 
\label{DD2.8}
D_{y}\psi_{i} = \frac{\psi_{i+1} - U_{i} \psi_{i}}{a},
\ee 
where 
\be 
\label{DD2.9}
U_{i}(x^{\mu}) = e^{iga A_{iy}(x^{\mu})}
\ee
is a ``link variable" (Wilson-line) to be introduced for gauge covariance, which connects the $i$ and $(i+1)$-th sites. $A_{iy}(x^{\mu})$ corresponds to the extra-space component of gauge field $A_{y}(x^{\mu}, y)$ in GHU. In order to guarantee the gauge covariance, the link variable should transform under the local gauge transformation as follows: 
\be 
\label{DD2.10}
U_{i} \ \ \to \ \ U'_{i} = g_{i+1}U_{i}g_{i}^{\dagger}, 
\ee 
where $g_{i}, \ g_{i+1}$ are group elements of $G_{i}, \ G_{i+1}$, respectively. Namely, $U_{i}$ behaves as bi-fundamental repr. of $(G_{i}, G_{i+1})$.   

We also need covariant derivative for a link variable $U_{i}$ in order to get the kinetic term for $A_{iy}$. 
Since $U_{i}$ behaves as the bi-fundamental repr. of $(G_{i}, G_{i+1})$, its 4D covariant derivative is given as 
\be 
\label{DD2.11}
D_{\mu}U_{i} 
=\partial_{\mu}U_{i} -igA_{i+1, \mu}U_{i} + ig U_{i}A_{i\mu}. 
\ee

\section{5D Scalar QED} 

As the simplest example to see the anomalous Higgs interaction, we take the model of 5D scalar QED on the latticized extra-space, following the prescription discussed in the previous section. 

The model is composed of a 5D scalar electron $\phi_{i}(x^{\mu})$ with electric charge $-e$ and the 5D photon \\ 
$(A_{i\mu}(x^{\mu}), A_{iy}(x^{\mu}))$. 
The lightest 4D field, corresponding to the KK zero-mode in GHU, of $A_{iy}(x^{\mu})$ is identified with the Higgs field and is supposed to have a VEV. Thus scalar electron has masses due to the VEV, though the gauge symmetry is not broken in this U(1) Abelian gauge theory. We expect in a realistic model with non-Abelian gauge symmetry to incorporate the standard model, gauge symmetry is broken through Hosotani-mechanism 
\cite{1983Hosotani}. 

The 4D lagrangian, which corresponds to the integral over $y$ of 5D lagrangian in GHU, is given by 
\be 
\label{DD3.1}
\mathcal{L} = a \sum_{i=1}^{N}\Bigg\{-\frac{1}{4}F_{i}^{\mu\nu}F_{i\mu\nu} 
+ \frac{1}{(ae)^{2}}(D^{\mu}U_{i})^\ast D_{\mu}U_{i} + (D^{\mu}\phi_{i})^\ast D_{\mu}\phi_{i} -(D_{y}\phi_{i})^\ast D_{y}\phi_{i} - m^2 \phi_{i}^\ast\phi_{i} \Bigg\}
\ee 
where
\begin{align}
&F^{\mu\nu}_{i}=\partial^{\mu}A_{i}^{\nu} - \partial^{\nu}A_{i}^{\mu},  
\label{DD3.2} \\ 
&U_{i} = e^{iaeA_{iy}},  
\label{DD3.3} 
\end{align}
and covariant derivatives are given as 
\begin{align}
&D^{\mu}U_{i} = \partial^{\mu}U_{i} -ieA_{i+1}^{\mu}U_{i} + ieU_{i} A_{i}^{\mu} \nonumber \\ 
&= i(ea) (\partial^{\mu}A_{iy} - \frac{A_{i+1}^{\mu} - A_{i}^{\mu}}{a})U_{i} 
= i(ea) (\partial^{\mu}A_{iy} - \partial_{y}A_{i}^{\mu})U_{i},  
\label{DD3.4} \\ 
&D^{\mu}\phi_{i} = \partial^{\mu}\phi_{i} + ieA_{i}^{\mu} \phi_{i},  
\label{DD3.5} \\ 
&D_{y}\phi_{i} = \frac{\phi_{i+1} - U_{i}^{\ast}\phi_{i}}{a}. 
\label{DD3.6} 
\end{align}
So far the charge $e$ and all fields are regarded to be 5D coupling and fields, respectively. We later introduce 4D electric charge $e_{4}$.   

\section{Kaluza-Klein modes and mass eigenvalues} 

\subsection{Kaluza-Klein mode expansion} 

We now perform ``discretized Fourier transform"  for each 5D field in order to get 4D mass eigenstates. 

First, let us note that although the translational invariance along the extra-space is violated by latticization, there still remain a symmetry in the theory under the following discrete transformation   
\be 
\label{DD4.1}
D: \ y_{i} \ \ \to \ \ y_{i+1}, \ \ \mbox{i.e.} \ \ \phi_{i} \ \ \to \ \ \phi_{i+1}, \ \mbox{etc.}
\ee
On the other hand, repeating $D$ $N$ times should be identity transformation. Thus, the eigenvalus of $D$ should be 
\be 
\label{DD4.2} 
(\omega_{N})^{n} \ \ (\omega_{N} \equiv e^{i\frac{2\pi}{N}}, \ n = 0, 1, 2, \cdots, N-1). 
\ee

Thus, the KK mode functions (vectors with $N$ elements) can be easily found without solving eingenvalue equation for 4D mass eigenvalues: 
\be 
\begin{pmatrix} 
\label{DD4.3}
1 \\ 
(\omega_{N})^{n} \\ 
(\omega_{N})^{2n} \\ 
\vdots \\ 
(\omega_{N})^{(N-1)n} \\ 
\end{pmatrix}.
\ee

By use of these eigenvectors we easily get (discretized) Fourier series expansion of each field,
\begin{align}
&A_{i \mu}(x^{\mu}) = \frac{1}{\sqrt{L}}\sum^{N-1}_{n = 0}A_{\mu}^{(n)}(x^{\mu})(\omega_{N})^{in}, 
\label{DD4.4} \\
&A_{iy}(x^{\mu}) = \frac{1}{\sqrt{L}}\sum^{N-1}_{n = 0}A_{y}^{(n)}(x^{\mu})(\omega_{N})^{in}, 
\label{DD4.5} \\ 
&\phi_{i}(x^{\mu}) = \frac{1}{\sqrt{L}}\sum^{N-1}_{n = 0} \phi^{(n)}(x^{\mu}) (\omega_{N})^{in},  
\label{DD4.6} 
\end{align}
where $\phi^{(n)}(x^{\mu})$ etc. are 4D fields with proper canonical mass dimension ($d = 1$) of KK mode $n$. 
The KK zero-mode of $A_{iy}$, $A_{y}^{(0)}$, is identified with the Higgs field. The reality of the gauge fields $A_{i\mu}, \ A_{iy}$ is guaranteed by  
\be 
\label{DD4.7} 
A_{\mu}^{(N-n)} = (A_{\mu}^{(n)})^{\ast}, \ \ \ A_{y}^{(N-n)} = (A_{y}^{(n)})^{\ast}. 
\ee 
Let us note that the KK zero-mode of $A_{i\mu}$, i.e. $A_{\mu}^{(0)}$ appears at each lattice site as $\frac{1}{\sqrt{L}}A_{\mu}^{(0)}(x^{\mu})$ (see (\ref{DD4.4})). Thus, the 4D electric charge $e_{4}$, which is nothing but the coupling constant of $A_{\mu}^{(0)}(x^{\mu})$ with the scalar electron is given by 
\be
\label{DD4.8}
e_{4}  = \frac{e}{\sqrt{L}} = \frac{e}{\sqrt{2\pi R}}, 
\ee
just as in the case of GHU. 

\subsection{4D mass eigenvalues} 

Getting mass eigenstates, we now calculate 4D mass eigenvalues of KK modes. 

\subsubsection{4D masses in the gauge-Higgs sector}

First we discuss the gauge-Higgs sector, i.e. the sector of $A_{\mu}^{(n)}$ and 
$A_{y}^{(n)}$. Note that this sector does not acquire the masses due to the VEV $v$ of the Higgs field $A_{y}^{(0)}$. 
We should also note that except for the Higgs field $A_{y}^{(0)}$, all non-zero KK modes of $A_{y}$ are absorbed by a sort of Higgs mechanism 
to the corresponding massive KK modes of $A_{\mu}$. 

Substituting the mode expansion (\ref{DD4.4}) and (\ref{DD4.5}) in the relevent part of the lagrangian (\ref{DD3.1}), 
\be 
\label{DD4.9}
a \sum_{i} \frac{1}{(ae)^{2}}(D^{\mu}U_{i})^\ast D_{\mu}U_{i}, 
\ee 
and performimg the sum over $i$ we get 
\be \label{DD4.10} 
\sum_{n=0}^{N-1} (\partial^{\mu}A_{y}^{(n)\ast} - \frac{(\omega_{N})^{-n}-1}{a}A_{\mu}^{(n)\ast})(\partial_{\mu}A_{y}^{(n)} - \frac{(\omega_{N})^{n}-1}{a}A_{\mu}^{(n)}).  
\ee
Here we have used the ortho-normal condition 
\be 
\label{DD4.11}
\sum_{i = 0}^{N-1}(\omega_{N})^{in} (\omega_{N})^{im} = N \delta_{n+m},    
\ee
where $n + m$ is in mod $N$. By the re-phasing of the fields $A_{\mu}^{(n)} \to -i(\omega_{N})^{-\frac{n}{2}}A_{\mu}^{(n)}$, the coefficients $\frac{(\omega_{N})^{n}-1}{a}$ is made real: 
\be 
\label{DD4.12}
\frac{(\omega_{N})^{n}-1}{a} \ \ \to \ \ -i \frac{(\omega_{N})^{\frac{n}{2}} - (\omega_{N})^{-\frac{n}{2}}}{a} = \frac{2\sin (\frac{n}{N}\pi)}{a}. 
\ee 
In this way it is clear that for the sector of non-zero KK modes Higgs-like mechanism is operative and the 4D mass eigenvalues of massive gauge bosons are given as 
\be 
\label{DD4.13} 
m_{n} = \frac{2\sin (\frac{n}{N}\pi)}{a}. 
\ee 
Note that in the ``continuum limit" $N \to \infty \ (a \to 0)$ keeping $L$ intact, the mass eigenvalue reduces to 
\be 
\label{DD4.14} 
m_{n} \to 2\frac{n}{Na}\pi = \frac{2n\pi}{L} = \frac{n}{R} \ \ (L = Na = 2\pi R), 
\ee 
which is nothing but the well-known KK masses in higher dimensional theories. 
These eigenvalues and the eigenvectors given in (\ref{DD4.3}) are just the same as those in the system of coupled harmonic oscillators: the system of springs and balls.  

\subsubsection{4D masses of matter field}  

Next, we discuss the mass eigenvalues of the scalar electron $\phi$. Again by substituting the mode expansion (\ref{DD4.6}) in the part relevant for the mass-squared term and replacing $A_{i}$ by its VEV, we get 
\be 
\label{DD4.15} 
- a \sum_{i} \{(D_{y}\phi_{i})^\ast D_{y}\phi_{i} + m^2 \phi_{i}^\ast \phi_{i}\}  
= - \sum_{n = 0}^{N-1} m_{n}^{2} |\phi^{(n)}|^{2},  
\ee 
where 
\be 
\label{DD4.16} 
m_{n}^{2} = \frac{1}{a^{2}}|(\omega_{N})^{n} - e^{-iae_{4} v}|^{2} + m^{2} = \{\frac{2}{a}\sin (\frac{n\pi}{N} + \frac{ae_{4}v}{2})\}^{2} + m^{2}. 
\ee
Again, at the continuum limit, the $m_{n}$ reduces to 
\be 
\label{DD4.17} 
m_{n}^{2} \to (\frac{n}{R} +e_{4}v)^{2} + m^{2},  
\ee
recovering the result in the GHU of 5D QED with $S^{1}$ compactification \cite{1998HIL}.

\section{The coupling constants of Higgs interaction} 

Our main purpose is to investigate whether the Higgs couplings with matter fields show some anomalous behaviour. 
In this section, we thus focus on the Higgs couplings with scalar electron, which is regarded as the counterpart of the Yuakawa coupling in a rearistic model, which can incorporate SM. 

We have obtained the mass-squared term for $n$-th KK mode of scalar electron $\phi$ (see (\ref{DD4.15}), \ (\ref{DD4.16})):  
\begin{equation} 
\label{DD5.1}
(\{\frac{2}{a}\sin (\frac{n\pi}{N} + \frac{ae_{4}v}{2})\}^{2} + m^{2}) \phi^{(n)}(x^{\mu})^{\ast}\phi^{(n)}(x^{\mu}).
\end{equation}
Since the physical Higgs field is nothing but the deviation of the Higgs field from its VEV, the Higgs interactions with $\phi$ are expected to be obtained by the following replacement in (\ref{DD5.1}): 
\be 
\label{DD5.2} 
v \ \ \to \ \ v + h(x^{\mu}) 
\ee
where $h$ denotes the physical Higgs. As has been already discussed in the introduction, in the case of 5D GHU the non-linearity of the mass eigenvalues came from the violation of translational symmetry along the extra-space due to the presence of the $Z_{2}$-odd bulk mass term for the fermion, and the Yukawa coupling was found to have ``off-diagonal" coulongs between diferrent KK modes \cite{2013HKLT}. Thus, the Higgs interactions obtained by the prescription in (\ref{DD5.2}) were argued to represent the ``diagonal" couplins between the same KK mode. It is interesting to note that in our model based on the scenario of DD, the non-linearlity comes from the fact that in the covariant derivative the Higgs is non-linearly realized from the begining. Thus Higgs interactions are expected to be diagonal in the base of KK modes, in contrast to the case of GHU. 

For instance, the coupling constant of 3-point coupling between Higgs and $n$-th KK mode of matter scalar $h\phi^{(n)\ast} \phi^{(n)}$ is given by the first 
derivative of the mass-squared $m_{n}^{2}$ in (\ref{DD5.1}), given by (\ref{DD4.16}), with respect to the VEV $v$, 
\be 
\frac{dm^2_n}{dv}
= \frac{2e_{4}}{a} \sin \left(\frac{2n\pi}{N} + ae_{4}v \right). 
\label{DD5.3} 
\ee

Similarly, the coupling constant of 4-point coupling $h^{2} \phi^{(n)\ast} \phi^{(n)}$ is calculated to be 
\be 
\frac{1}{2}\frac{d^{2}m^2_n}{dv^{2}}
= e_{4}^{2} \cos \left(\frac{2n\pi}{N} + ae_{4}v \right). 
\label{DD5.4} 
\ee

Especially the coupling constants with KK zero-mode of matter scalar are given as 
\bea 
&& \mbox{3-point coupling}: \ \ \frac{2e_{4}}{a} \sin \left(ae_{4}v \right),  
\label{DD5.5} \\ 
&& \mbox{4-point coupling}: \ \ e_{4}^{2} \cos \left(ae_{4}v \right). 
\label{DD5.6}
\eea
We realize that, for instance, the 4-point coupling (\ref{DD5.6}) shows quite different behaviour from that of ordinary 4D theory such as SM, 
and even vanishes for a specific value of the VEV, 
\be 
\label{DD5.7} 
v = \frac{\pi}{2ae_{4}},   
\ee
just as in the case of Yukawa coupling in 5D GHU mentioned in the introduction \cite{2013HKLT}, \cite{2007HS}, \cite{2008HOOS}, \cite{2009HK}, \cite{2009HKT}.

\section{Anomalous Higgs interaction}

In this section we finally discuss how our predictions on the Higgs couplings deviate from the corresponding predictions in the SM. 

First we discuss what interaction in this model should be compared with the Yukawa coupling in the SM and then we compare our prediction on that observable with that of the SM.

\subsection{What is the counterpart of the Yukawa coupling in the SM ?} 

Our model, latticized 5D scalar QED, is just a toy model and is not realistic, in particular in the following sense: \\ 
$\cdot$ Gauge group is Abelian U(1) and does not contain the gauge symmetry SU(2)$\times$U(1) of the SM (even if we restrict ourselves to the electro-weak sector). \\ 
$\cdot$ The matter field is a scalar, not fermions to describe quarks and leptons. \\ 
One reason not to have introduced fermions is to avoid the cumbersome problem which arises when we formulate the theory of fermions on a lattice. Although we believe that the anamalous Higgs interaction pointed out in this work has its origin in the fact that the extra dimension is latticized and therefore is a general feature of the latticized higher dimensional gauge theories, when we wish to compare our prediction with that of SM it is not trivial what interaction 
in our model should be counterpart of the interaction in the SM we are familiar with, such as the Yukawa coupling. 

From such a point of view, it may be useful to think of the situation in the supersymmetric theory for our reference, where the relation between the coupling constants of matter fermion and matter scalar becomes clear, though we do not intend to introduce supersymmetry to our theory. In MSSM, for instance, a term in the superpotential 
\be 
\label{DD5.8} 
W = fh\bar{t}t + \cdots  
\ee 
provides not only the Yukawa coupling coustant $f$ for the top quark, but also 4-point interaction between the Higgs and stop $\tilde{t}$, the super-partner of $t$, $h^{2} |\tilde{t}|^{2}$ with a coupling constant $f^{2}$:   
\be 
\label{DD5.9}
fh\bar{t}t
\stackrel{\mathrm{SUSY}}{\Longleftrightarrow}f^2 h^{2}\tilde{t}^{\ast}\tilde{t}.
\ee

This argument suggests that the counterpart of the Yukawa couling in our model is the coupling constant of 4-point function given in (\ref{DD5.6}), which we write as 
$f_{\mathrm{DD}}^{2}$ from now on, rather that the coupling constant of 3-point function (\ref{DD5.5}): 
\be 
\label{DD5.10}
f_{\mathrm{DD}}^{2} \equiv e_{4}^{2} \cos \left(ae_{4}v \right).   
\ee
  
\subsection{Quantitative analysis of the anomaly} 

We now compare the prediction of our model $f_{\mathrm{DD}}^{2}$ given in (\ref{DD5.10}) with the squared Yukawa coupling 
in the SM, $f_{\mathrm{SM}}^{2}$. In this analysis we set $m = 0$. This seems to be reasonable, since in a realistic theory, 
ordinary matter fields are expected to have masses only through spontaneous gauge symmetry breaking. 

In the standard model, the Yukawa coupling is simply given by the ratio of the frmion mass, given by $m^{(0)}$ in our model, to the VEV of the Higgs. Hence we identify the Yukawa coupling of the SM as 
\be 
\label{DD5.11}
f_{\mathrm{SM}} = \frac{m^{(0)}}{v_{\mathrm{SM}}}.  
\ee 
One non-trivial thing here is whether the $v_{\mathrm{SM}}$ is identical with $v$ in our model. Though the VEV is determined so that it 
provides corrcet weak scale $M_{W}$, $M_{W}$ may not be linear in $v$ in our model in contrast to the case of the SM where $M_{W}$ is linear in $v_{\mathrm{SM}}$, $M_{W} = \frac{g}{2}v_{\mathrm{SM}}$. We have to wait until a realistic model incorporating the SM is constructed in order to make a definite claim on this issue. So now we assume two typical possible cases and discuss the anomalous Higgs interaction in these two cases separately. 

\subsubsection{Case 1: $v_{\mathrm{SM}} = v$} 

We first consider the case where $M_{W}$ is linear function of the VEV in the DD scenario as well: $M_{W} = \frac{g}{2}v$, i.e. 
\be 
\label{DD5.12} 
v_{\mathrm{SM}} = v. 
\ee
Then from (\ref{DD5.11}) $f_{\mathrm{SM}} = \frac{m^{(0)}}{v}$ and therfore we get from (\ref{DD5.10}) and $m^{(0)} = \frac{2}{a}\sin (\frac{ae_{4}v}{2})$ (see (\ref{DD4.16}) for $n = 0, \ m = 0$)       
\be 
\label{DD5.13} 
\frac{f_{\mathrm{DD}}^{2}}{f_{\mathrm{SM}}^{2}} = \frac{e_{4}^{2}v^{2} \cos \left(ae_{4}v \right)}{\{\frac{2}{a}\sin (\frac{ae_{4}v}{2})\}^{2}} = x^{2}(\cot^{2}x - 1)
\ee
where the dimensionless parameter $x$ is defined as 
\be 
\label{DD5.14}
x \equiv \frac{ae_{4}v}{2}. 
\ee
For $x > 0$ the ratio in (\ref{DD5.13}) deviates from unity and the Yukawa coupling (its counterpart) becomes anomalous. 

\subsubsection{Case 2: $v_{\mathrm{SM}} \neq v$} 

Since the violation of the translational invariance is due to the property of the extra space, we may naturally expect that in a realistic model the gauge boson mass $M_{W}$ is also given by trigonometric funcition of $v$ like that in (\ref{DD4.16}) with $n = 0$ and $m = 0$:
\be 
\label{DD5.15} 
M_{W} = \frac{2}{a}\sin (\frac{agv}{4}),  
\ee
which reduces to $\frac{g}{2}v$ in the limit $a \to 0$. This means  
\be 
\label{DD5.16}
v_{\mathrm{SM}} = \frac{4}{ag}\sin (\frac{agv}{4}). 
\ee
Since $e_{4}$ in our toy model should be identified with $\frac{g}{2}$ in a realistic model, we replace $\frac{g}{2}$ in (\ref{DD5.16}) by $e_{4}$ to get 
\be 
\label{DD5.17}
v_{\mathrm{SM}} = \frac{2}{ae_{4}}\sin (\frac{ae_{4}v}{2}). 
\ee
Thus we get a simple result,  
\be 
\label{DD5.18} 
f_{\mathrm{SM}} = \frac{m^{(0)}}{v_{\mathrm{SM}}} = e_{4}, 
\ee
and therefore 
\be 
\label{DD5.19} 
\frac{f_{\mathrm{DD}}^{2}}{f_{\mathrm{SM}}^{2}} = \cos (ae_{4}v) = \cos (2x). 
\ee 
Again, for $x > 0$ the ratio in (\ref{DD5.19}) deviates from unity and the Yukawa coupling (its counterpart) becomes anomalous.

\subsubsection{Continuum limit and decoupling limit}   

We finally consider two physically interesting limits \\ 
(i) Continuum limit: \ \ $N \to \infty, \ \ a \to 0$ keeping $L = Na = 2\pi R$ intact. \\ 
(ii) ``Decoupling limit": \ \ $\frac{M_{W}}{M_{c}} \to 0$ keeping $N$ as a finite integer \ ($M_{c} \equiv \frac{1}{R}$). \\ 
Clearly (i) is the limit where original 5D GHU with $S^{1}$ compactification \cite{1998HIL} is recovered. Since there is no other source of the violation of the translational invariance except for the latticization, we expect that the SM prediction is recovered in this continuum limit. (ii) is a limit where 
the masses of all non-zero KK modes of the order of $M_{c}$ (compactification mass scale) are much greater than the weak scale 
and these massive KK particles are expected to decouple from the low energy effetive theory, thus recovering the SM. In our toy model 
$M_{W}$ should be regarded as $\sim e_{4}v$ and the decoupling limit is equivalent to $e_{4} v R \sim e_{4} v L = e_{4} v Na \sim e_{4}v a \to 0$. 

It is now easy to know that in both limits $x = \frac{ae_{4}v}{2} \to 0$. Interestingly, in both cases of (\ref{DD5.13}) and (\ref{DD5.19}), in this limit the ratio $\frac{f_{\mathrm{DD}}}{f_{\mathrm{SM}}} \to 1$ and the anomaly just goes away, as we expected. 

\section{Summary} 

In this paper we discussed anomalous Higgs interaction with matter field in the scenario of dimensional deconstruction (DD) \cite{2001ACG}, which can accommodate light Higgs suggested by recent LHC experiments. In the snenario Higgs is regarded as a pseudo-Nambu-Goldstoe boson (PNGB). Though the scenario is formulated in ordinary 4-dimensional (4D) space-time, the PNGB may be interpreted as the Kaluza-Klein (KK) zero-mode of the extra-space component of higher dimensional gauge field. Namely, the scenario of dimensional deconstruction may be interpreted as ``latticized" version of gauge-Higgs unification (GHU) \cite{1979Manton}, \cite{1983Hosotani}, \cite{1998HIL}. 

In our previous paper discussing anomalous Higgs interaction in GHU \cite{2013HKLT}, we pointed out that the interplay between the periodicity of physical observables in the Higgs field and the violation of translational invariance along the extra-space plays an essential role to get the anomalous Higgs interaction. We argued that in the scenario of DD, the interplay between two ingredients is also present. Namely, the periodicity in the Higgs field is guaranteed by the fact that Higgs field is non-linearly realized in the form of link-variable connecting neighboring lattice sites and the violation of translational invariance is realized by the fact that extra-space is latticized, when DD is regarded as a sort of GHU. 

We took the attitude that DD is latticized GHU, and adopted 5-dimensional scalar QED with scalar electron as its matter field, as a toy model. 

We discussed the dimensionless coupling constant of 4-point interaction of Higgs with scalar electron, $h^{2}\phi^{\ast}\phi$ ($h$: Higgs, $\phi$: scalar electron), as the counterpart of Yukawa coupling in a realistic model with matter fermion. By use of the calculated (4D) mass eigenvalue for the KK zero-mode of the matter field, we derived the dimensionless coupling constant and have confirmed that the coupling is anomalous, i.e. that it deviates from what we expect in the standard model. 

Though the derived anamalous interaction has much similarity to the one in the GHU \cite{2013HKLT}, \cite{2007HS}, \cite{2008HOOS}, \cite{2009HK}, \cite{2009HKT}, the anomalous interaction in DD scenario has its own characteristic feature, which is not shared by the GHU scenario. Namely, the anomaly exists even though we do not introduce ``$Z_{2}$-odd bulk mass" for fermion as is found in the case of GHU with orbifold as its extra-space. This is because the violation of translational invariance is not due to the presence of the bulk mass term, but due to the fact that the extra-space itself, on which theory is constructed, is latticized. Thus the anamalous interactions are expected to appear not only at the sector of matter scalar but in all sectors of the theory. On the other hand, it means that anomaly goes away (unless we do not introduce another source of the violation of translational invariance such as the bulk mass term) in the continuum limit of lattice spacing $a \to 0$. We have confirmed this property by explicit calculation of the anomalous 
coupling. 

The model we adopted is just a toy model, since our main purpose was to show explicitly that the expected anomalous Higgs interaction really exists. To make the theory realistic we have to achieve several things; In order to incorporate the SM, gauge group should be enlarged to non-Abelian symmetry such as SU(3) and matter fermion also should be introduced, though to put fermion on the lattice space causes some cumbersome problem in general. We also have to think about how we can realize the orbifold and $Z_{2}$-odd bulk mass for chiral fermion on the lattice. 

In spite of these remaining issues, the scenaio of DD is attractive since it is renormalizable 4D gauge theory and 
realizes at the same time the mechanism of GHU to solve the hierarachy problem \cite{1998HIL}, by replacing the sum over infinite number of KK modes in the intermediate state of the quantum correction to the Higgs mass by the sum over just $N$ KK modes ($N$: the number of the lattice sites).   
  
Finally, we point out that another interesting scenario of BSM, closely related to DD and GHU, i.e. Little Higgs probably also possesses anomalous Higgs interaction. At least, the non-linearity in the Higgs field should exist, since Higgs is regarded to be PNGB and therefore non-linearly realized, just as in the scenario of DD (see (\ref{DD1.1})).     
     
%%%%%%%%%%%%%%%%%%%%%%%%%%%%%%%%%%%%%%%%%%%%%%%%%%%%%%%%%% 
%%                 %%%%%%%%%%%%%%%%%%%%%%%%%%%%%%%%%%%%%%%
%% Acknowledgments %%%%%%%%%%%%%%%%%%%%%%%%%%%%%%%%%%%%%%%
%%                 %%%%%%%%%%%%%%%%%%%%%%%%%%%%%%%%%%%%%%%
%%%%%%%%%%%%%%%%%%%%%%%%%%%%%%%%%%%%%%%%%%%%%%%%%%%%%%%%%%
\subsection*{Acknowledgments}
This work of C.S.L. was supported by the Grant-in-Aid for Scientific Research 
of the Ministry of Education, Culture, Sports, Science and Technology, Nos.~21244036,~23654090,~23104009. 

%%%%%%%%%%%%%%%%%%%%%%%%%%%%%%%%%%%%%%%%%%%%%%%%%%%%%%%%%%
%%            %%%%%%%%%%%%%%%%%%%%%%%%%%%%%%%%%%%%%%%%%%%%
%% Bliography %%%%%%%%%%%%%%%%%%%%%%%%%%%%%%%%%%%%%%%%%%%%
%%            %%%%%%%%%%%%%%%%%%%%%%%%%%%%%%%%%%%%%%%%%%%%
%%%%%%%%%%%%%%%%%%%%%%%%%%%%%%%%%%%%%%%%%%%%%%%%%%%%%%%%%%

\end{document}